\documentclass[aps,prl,twocolumn,reprint,floats,epsfig,showpacs,groupedaddress,superscriptaddress]{revtex4}

\usepackage{bbm}
\usepackage{amssymb}
\usepackage{amsmath}
\usepackage{ifpdf}
\usepackage{color}
\usepackage{graphicx,epsfig}
\usepackage{hyperref}

\newcommand{\1}{{\mathbbm{1}}}

\graphicspath{
{./}
{./Figures/}
}

\begin{document}

\title{Real-Time Simulation of Non-Equilibrium Transport of Magnetization \\
in Large Open Quantum Spin Systems driven by Dissipation}

\author{D. Banerjee}
 \affiliation{NIC, DESY Zeuthen, Platanenallee 6, 15738 Zeuthen, Germany}
\author{F. Hebenstreit}
 \affiliation{Albert Einstein Center, Institute for Theoretical Physics, Bern University, 3012 Bern, Switzerland}
\author{F.-J. Jiang}
 \affiliation{Department of Physics, National Taiwan Normal University, 88, Sec.~4, Ting-Chou Rd., Taipei 116, Taiwan}
\author{U.-J. Wiese}
 \affiliation{Albert Einstein Center, Institute for Theoretical Physics, Bern University, 3012 Bern, Switzerland}

\begin{abstract}
Using quantum Monte Carlo, we study the non-equilibrium transport of 
magnetization in large open strongly correlated quantum spin $\frac{1}{2}$ 
systems driven by purely dissipative processes that conserve the uniform or 
staggered magnetization, disregarding unitary Hamiltonian dynamics.
We prepare both a low-temperature Heisenberg 
ferromagnet and an antiferromagnet in two parts of the system that are initially
isolated from each other. We then bring the two subsystems in contact and study 
their real-time dissipative dynamics for different geometries. The flow of the
uniform or staggered magnetization from one part of the system to the other is 
described by a diffusion equation that can be derived analytically.
\end{abstract}

\maketitle

Simulating the real-time evolution of large strongly correlated quantum systems 
is notoriously difficult, due to the dimension of the Hilbert space, which 
grows exponentially with the system size. In this case, Monte Carlo methods are 
usually not applicable because importance sampling is prevented by severe sign 
or complex phase problems \cite{Tro05}. While in Euclidean time some severe 
sign problems have been solved using the meron-cluster algorithm 
\cite{Bie95,Cha99} or the fermion bag method \cite{Cha10,Cha12,Cha13}, until 
recently real-time simulations of quantum systems have been limited to small 
volumes that are accessible to exact diagonalization, or to gapped 1-d systems 
to which the time-dependent density matrix renormalization group (DMRG)  
\cite{Whi92,Sch05} can be applied. Even then, due to the growth of 
entanglement, only moderate time intervals can be investigated 
\cite{Vid03,Whi04,Ver04,Zwo04,Dal04,Bar09,Piz13}. Dynamical phenomena in 
non-equilibrium quantum systems have been studied in 
\cite{Cor74,Ber02,Aar02,Ber08,Die08,Dal09,Die10,Mue12,Sie13,DeG13,Hor13,Chi15}. 
Recently, we have developed a new Monte Carlo method that allows us to simulate 
the real-time evolution of large strongly coupled quantum systems in any 
dimension for an arbitrary amount of time, for specific dynamics driven by 
purely dissipative processes that are described by a Lindblad equation 
\cite{Ban14,Heb15}. In particular, the unitary time-evolution driven by a 
Hamiltonian, which would give rise to a severe complex phase problem, has been 
replaced by a dissipative process. Still severe sign problems arise even for 
the purely dissipative dynamics, but they have been solved analytically by 
identifying exact cancellations in the corresponding real-time path integral. 
Purely dissipative processes play an important role in quantum information 
processing, for example, in order to prepare specific states for quantum 
computation \cite{Rau01,Nie01,Chi02,Ali04,Ver09} or entanglement generation
\cite{Kra11}. The control of quantum systems by measurements has been
investigated in \cite{Sug05,Pec06}. Ultracold atoms in optical lattices or 
trapped ions provide platforms in which such dynamics can be engineered in
quantum simulation experiments \cite{Cir12,Lew12,Blo12,Bla12}. 

In this paper, our primary goal is not yet to make contact with concrete cold
atoms experiments. Instead, we demonstrate that our ability to classically
simulate the real-time dynamics of engineered dissipative processes in large 
open quantum spin systems puts us in a unique position to study transport 
phenomena far away from equilibrium. Such processes thus provide a bridge between
classical and quantum simulations of real-time quantum dynamics. Here we 
investigate a low-temperature Heisenberg ferromagnet and an antiferromagnet 
which are initially isolated from each other in two separate parts of the 
volume. The two parts, which act as large reservoirs of uniform or staggered 
magnetization, are then put in contact and evolve in time according to a 
dissipative process which either conserves the uniform or the staggered 
magnetization. The corresponding conserved quantity then flows from its 
reservoir into the other half of the system, through an opening whose size we 
vary. The non-equilibrium diffusive processes are driven by the gradient of the 
corresponding conserved quantity. They come to an end only when the staggered 
or uniform magnetization is homogeneously distributed throughout the entire 
system. Remarkably, certain aspects of the dynamics are described by a 
classical diffusion equation which can be derived analytically from the 
underlying dissipative quantum dynamics. Significantly extdending 
previous work \cite{Ban14,Heb15}, the current setting allows us to study the 
diffusion process of the conserved quantity in real-space.

We consider systems of quantum spins $\frac{1}{2}$ on a square lattice, which
are dissipatively coupled to their environment. The dynamics is characterized 
by a set of Lindblad operators \cite{Kos72,Lin76,Kra83}, $L_{o_k}$, that obey 
$(1 - \varepsilon \gamma) \1 + \sum_{k,o_k} L_{o_k}^\dagger L_{o_k} = \1$, where
$\varepsilon$ is a small time-step. The Lindblad operators induce quantum jumps 
and $\gamma$ determines their probability per unit time. We will 
analytically derive the relation between the parameter $\gamma$ and the diffusion 
coefficient of the classical diffusion equation. The time-evolution of the density 
matrix is then determined by the Lindblad equation
\begin{equation}
\label{Lindblad}
\partial_t \rho = \frac{1}{\varepsilon} \sum_{k,o_k} \left(
L_{o_k} \rho L_{o_k}^\dagger - \frac{1}{2} L_{o_k}^\dagger L_{o_k} \rho - 
\frac{1}{2} \rho  L_{o_k}^\dagger L_{o_k} \right) \, . 
\end{equation}

We will consider two different dissipative processes whose jump operators 
$L_{o_k} = \sqrt{\varepsilon \gamma} P_{o_k}$ are determined by operators $P_{o_k}$
that project on the eigenstates of an observable $O$ with eigenvalue $o_k$. For 
the first process (process 1), which conserves the uniform magnetization vector,
the observable is the total spin $O^{(1)} = (\vec S_x + \vec S_y)^2$ of a pair of
spins $\vec S_x$ and $\vec S_y$ located on neighboring lattice sites $x$ and 
$y$. The projection operators corresponding to total spin 1 or 0 are then given 
by 
\begin{equation}
P_1 = \left(\begin{array}{cccc} 
1 & 0 & 0 & 0 \\ 0 & \frac{1}{2} & \frac{1}{2} & 0 \\
0 & \frac{1}{2} & \frac{1}{2} & 0 \\ 0 & 0 & 0 & 1 \end{array}\right), \quad
P_0 = \left(\begin{array}{cccc} 
0 & 0 & 0 & 0 \\ 0 & \frac{1}{2} & - \frac{1}{2} & 0 \\
0 & - \frac{1}{2} & \frac{1}{2} & 0 \\ 0 & 0 & 0 & 0 \end{array}\right).
\end{equation}
As we have shown in \cite{Ban14,Heb15}, the conservation of the total spin in
this dissipative process implies that the low-momentum modes of the 
magnetization equilibrate very slowly. The second dissipative process (process 
2), which conserves the 3-component of the staggered magnetization, is 
characterized by the observable $O^{(2)} = S_x^+ S_y^+ + S_x^- S_y^-$ with the 
three projection operators
\begin{eqnarray}
&&P_+ = \left(\begin{array}{cccc} 
\frac{1}{2} & 0 & 0 & \frac{1}{2} \\ 0 & 0 & 0 & 0 \\ 0 & 0 & 0 & 0 \\ 
\frac{1}{2} & 0 & 0 & \frac{1}{2} \end{array}\right), \quad
P_0 = \left(\begin{array}{cccc} 
0 & 0 & 0 & 0 \\ 0 & 1 & 0 & 0 \\ 0 & 0 & 1 & 0 \\ 0 & 0 & 0 & 0 
\end{array}\right), \nonumber \\
&&P_- = \left(\begin{array}{cccc} 
\frac{1}{2} & 0 & 0 & - \frac{1}{2} \\ 0 & 0 & 0 & 0 \\ 0 & 0 & 0 & 0 \\ 
- \frac{1}{2} & 0 & 0 & \frac{1}{2} \end{array}\right). 
\end{eqnarray}
In this case, as we showed in \cite{Heb15}, the high-momentum modes of the
magnetization (namely those with momenta near the conserved $(\pi,\pi$)-mode 
representing the staggered magnetization) equilibrate very slowly. Both
dissipative processes ultimately converge to a trivial infinite-temperature
density matrix that is proportional to the unit-matrix, at least within the
sector defined by the value of the corresponding conserved quantity.

As discussed in detail in \cite{Ban14,Heb15}, the Lindblad equation can be
represented by a path integral consisting of a Euclidean time contour that 
defines an initial density matrix in thermal equilibrium and a real-time 
Schwinger-Keldysh contour \cite{Sch61,Kel65} that leads from an initial time 
$t_0$ to a final time $t_f$ and back. Remarkably, the probability to reach a 
specific final state $|f\rangle$ can be computed very efficiently with a 
loop-cluster algorithm, similar to the one used in Euclidean time 
\cite{Eve93,Wie94,Bea96}. The cluster rules have been discussed in detail in
\cite{Heb15}.

\begin{figure}[tbp]
\includegraphics[width=0.43\textwidth]{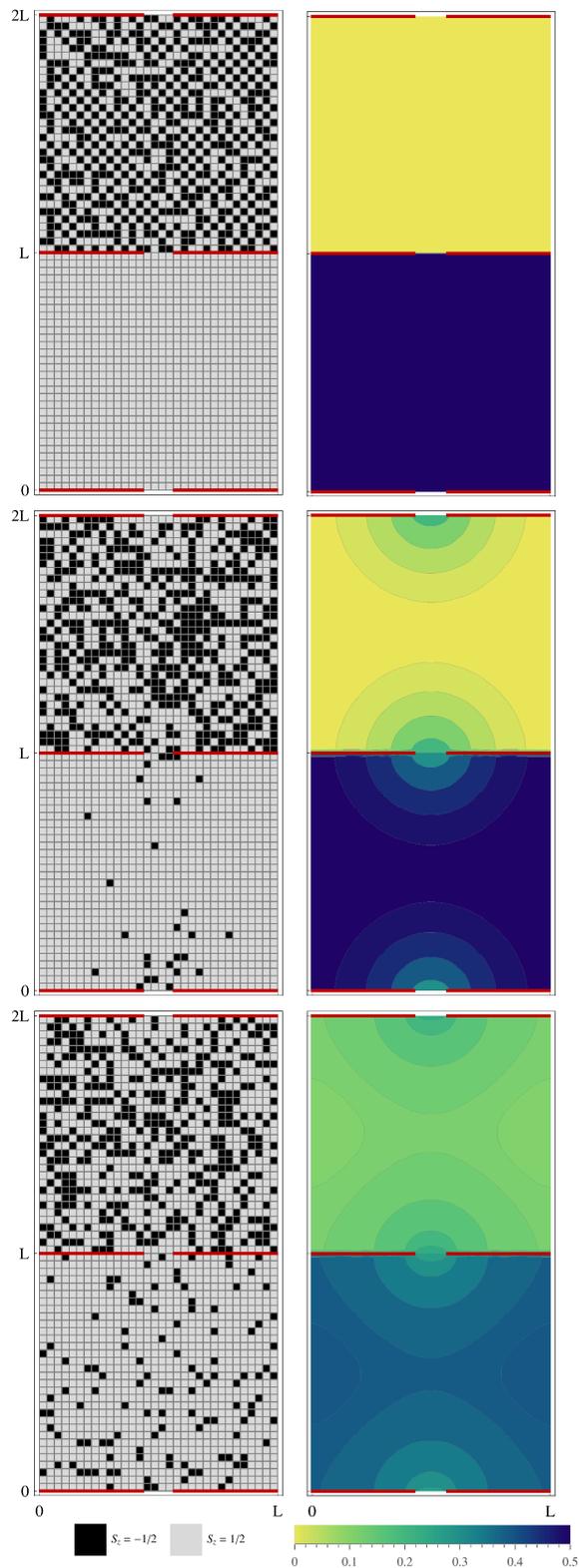}
\caption{[Color online] 
\textit{Real-time evolution of the uniform \mbox{magnetization} on a $32 \times 64$ lattice with an opening of size $L' = 4a$ for a total uniform magnetization value $M_u=\frac{1}{2}(L/a)^2=512$ at initial temperature $\beta J=80$. 
Typical configurations (left) and expectation values of the uniform magnetization (right) at time $t = 0$ (top), $50/\gamma$ (middle), and $500/\gamma$ (bottom).}}
\label{uniform}
\end{figure}

\begin{figure}[tbp]
\includegraphics[width=0.43\textwidth]{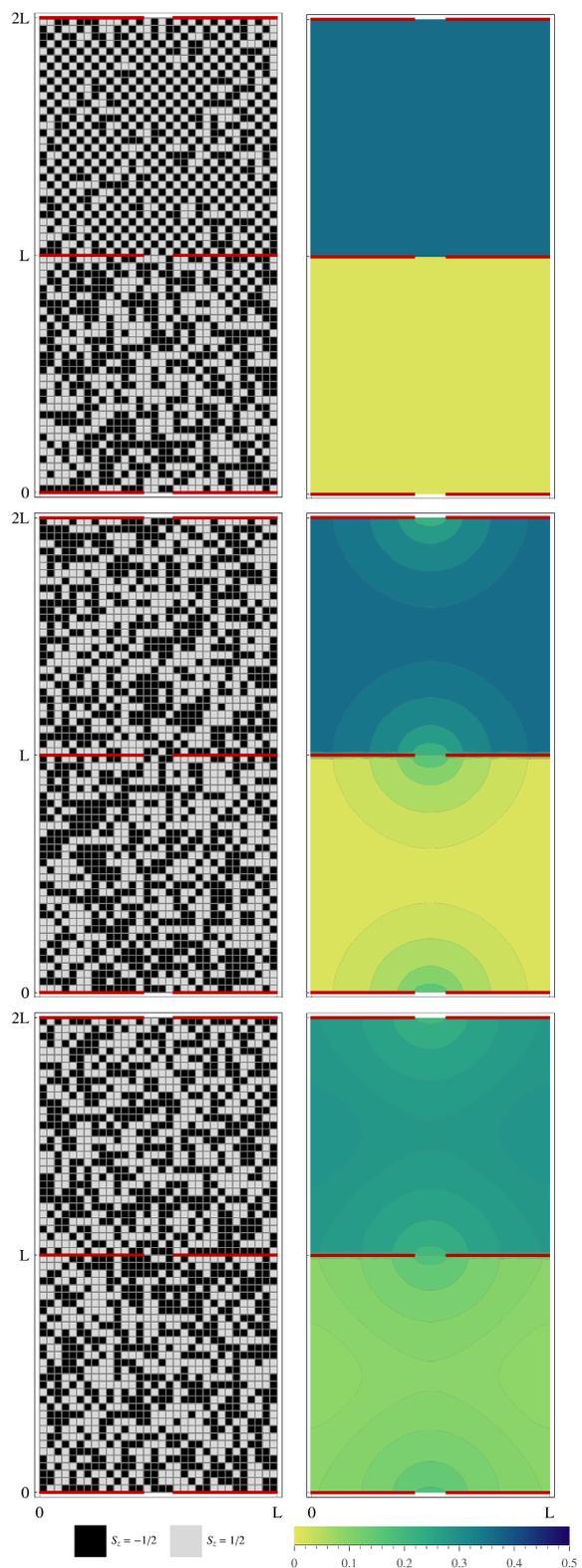}
\caption{[Color online] 
\textit{Real-time evolution of the staggered magnetization on a $32 \times 64$ lattice with an opening of size $L' = 4a$ for a total staggered magnetization value $M_s=\frac{3}{8}(L/a)^2=384$ at initial temperature $\beta J = 80$. 
Typical configurations (left) and expectation values of the staggered magnetization (right) at time $t = 0$ (top), $50/\gamma$ (middle), and $500/\gamma$ (bottom).}}
\label{staggered}
\end{figure}

\begin{figure}[tbp]
\includegraphics[width=0.45\textwidth]{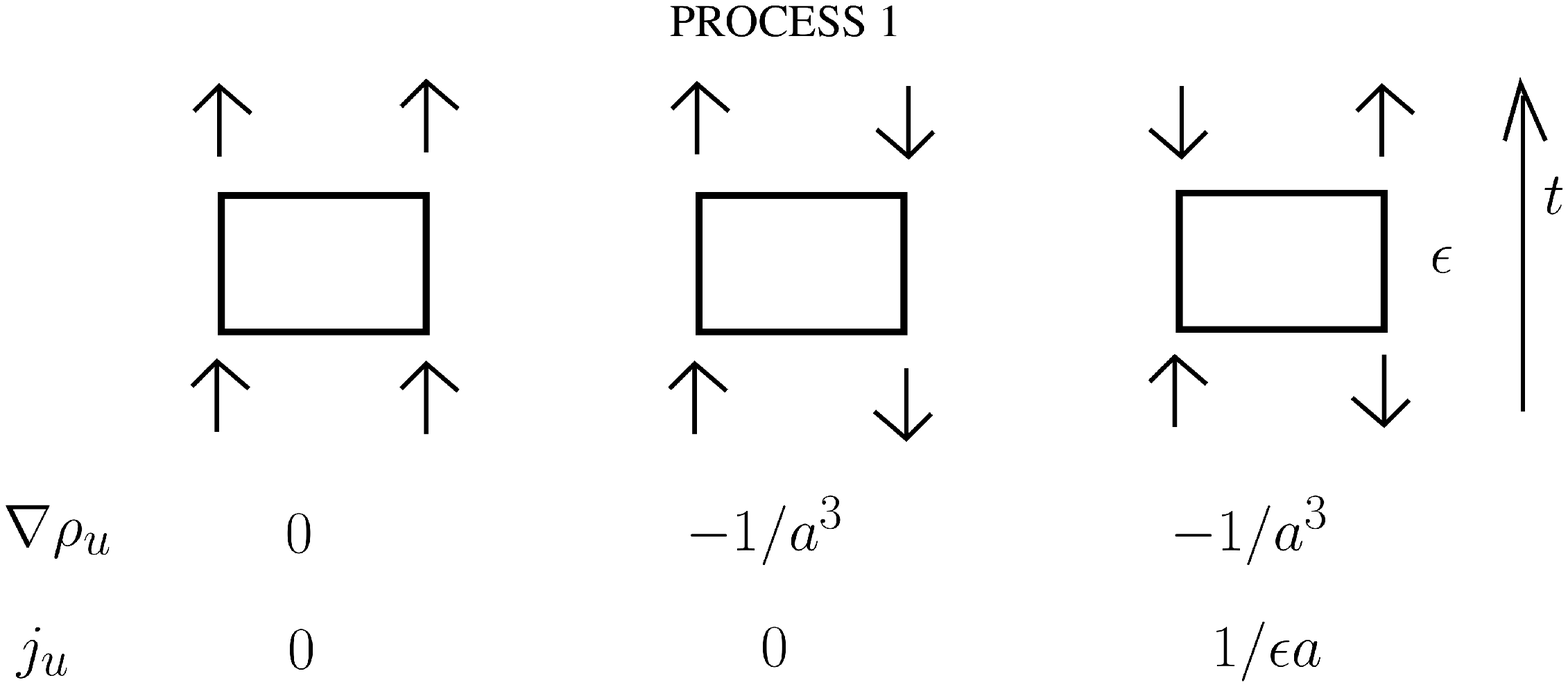}
\includegraphics[width=0.45\textwidth]{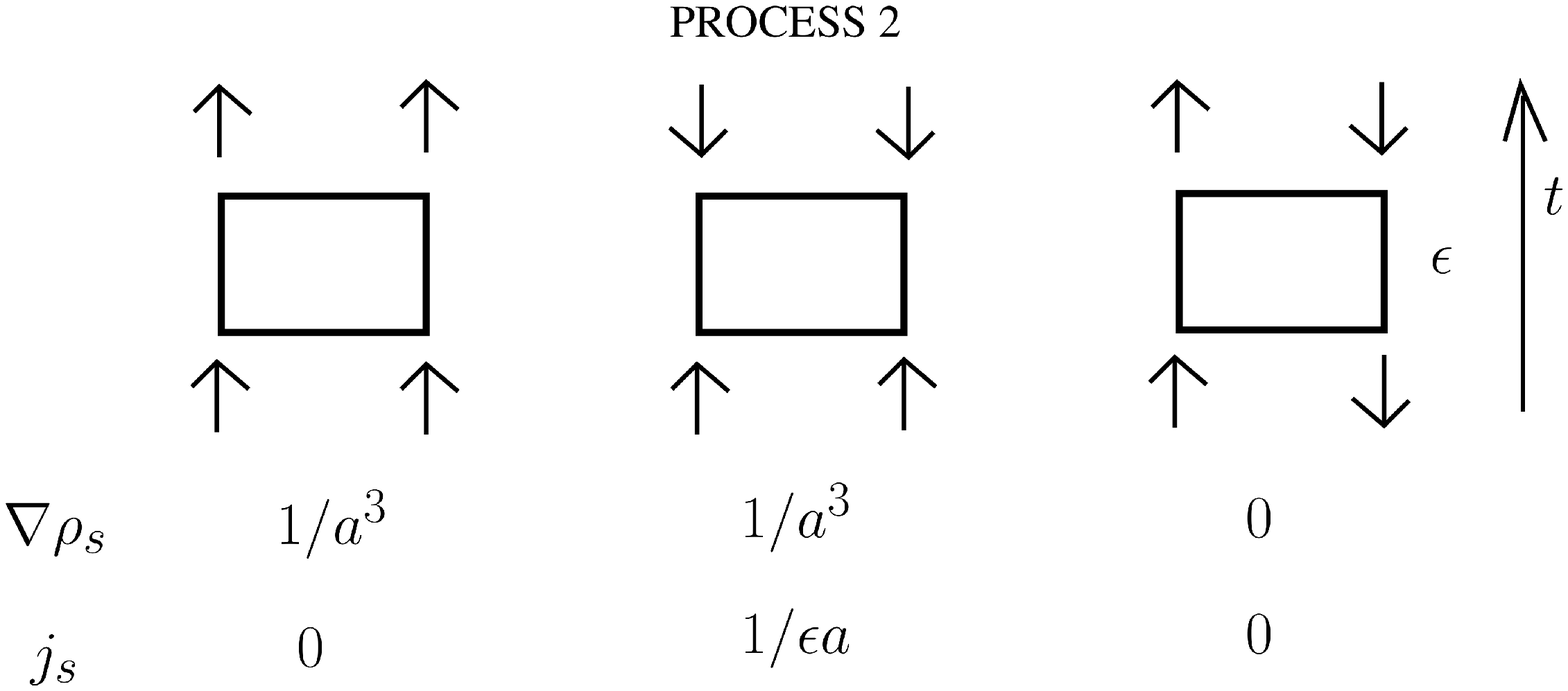}
\caption{\textit{Configurations of two neighboring spins evolving
in time, together with the resulting values for $\rho_x(t)$ and $j_{x,i}(t)$ for
the dissipative process 1 and 2, that conserves the uniform and staggered
magnetization, respectively. The current is driven by the gradient of the
corresponding density.}}
\label{configs}
\end{figure}

We consider a spin $\frac{1}{2}$ Heisenberg model with Hamiltonian 
$H\! =\! J \sum_{\langle xy \rangle}\! \vec S_x \cdot \vec S_y$. In order to prepare an 
initial density matrix we consider an $L\times2L$ lattice that is divided 
into two subsystems of size $L \times L$ each with individual periodic boundary 
conditions. One system is antiferromagnetic (with $J\!>\!0$) and the other is
ferromagnetic and has the opposite exchange coupling. Both subsystems are
initialized at the same temperature. The initial density matrix is then 
subjected to one of the two dissipative real-time processes. During the 
real-time process the two subsystems are put in contact through two openings of
size $L' \leq L$ on opposite sides of both systems. This is achieved by changing
the original boundary conditions with period $L$ on two sets of $L'$ links. 
These links connect the two subsystems, so that the total system now has 
boundary conditions with period $2 L$ in a strip of transverse size $L'$ and the
original pair of boundary conditions with period $L$ on the remaining strip of
transverse size $L - L'$. The transverse direction of size $L$ always has 
ordinary periodic boundary conditions. Using the loop-cluster algorithm we 
calculate the expectation value of the 3-component for each spin $S_x^3$ at the 
time $t_f$ when the 3-components of all spins are finally being measured.
The data are separately analyzed for each total value of the conserved uniform 
or staggered magnetization. By using an improved estimator similar to the one
constructed in \cite{Ger09,Ger11}, we increase the statistics by a factor that
grows exponentially with the number of loop-clusters. This improves the 
accuracy of the numerical data very substantially and leads to the results
depicted in Fig.~\ref{uniform} (uniform magnetization) and Fig.~\ref{staggered}
(staggered magnetization). As we have discussed in detail in \cite{Ban14,Heb15},
the dissipative processes give rise to different time scales. While process 1
quickly destroys the initial antiferromagnetic order over a time scale 
$1/\gamma$, the conserved uniform magnetization undergoes a much slower 
diffusion process. In particular, in process 1 the magnetization modes with low 
momentum $p$ equilibrate only over time scales $1/(\gamma a^2 p^2)$, where $a$
is the lattice spacing. Similarly, in process 2, which conserves the staggered 
magnetization, the modes with momenta near $(\pi,\pi)$ are severely slowed down. 
The dissipative dynamics can be characterized as a heating process that affects 
different modes at different time scales.
While the underlying diffusive processes are quantum mechanical, the resulting
expectation values of the conserved uniform or staggered magnetization are
described by a classical diffusion equation 
\begin{equation}
\label{diffusion}
\partial_t \rho_x(t) = 
\frac{\gamma}{2} 
\sum_i \left[\rho_{x+a\hat i}(t) - 2 \rho_x(t) + \rho_{x-a\hat i}(t)\right] \, . 
\end{equation}
Here $\rho_x(t)$ is the expectation value of the conserved quantity at the 
lattice site $x$ at time $t$ and $\hat i$ is the unit-vector in the 
$i$-direction. Interestingly, the classical diffusion equation can be derived 
analytically from the underlying quantum spin dynamics, and the diffusion 
coefficient is determined by the parameter $\gamma$ that drives the 
Lindblad process of eq.~\eqref{Lindblad}. The lattice diffusion equation 
\eqref{diffusion} results from the continuity equation \vspace{-0.1cm}
\begin{equation}
\label{continuity}
\partial_t \rho_x(t) + 
\frac{1}{a} \sum_i \left[j_{x,i}(t) - j_{x-a\hat i,i}(t)\right] = 0 \, , \vspace{-0.1cm}
\end{equation}
combined with the lattice gradient equation \vspace{-0.1cm}
\begin{equation}
\label{gradient}
j_{x,i}(t) = - \frac{a\gamma}{2} \left[\rho_{x+a\hat i}(t) - \rho_x(t)\right] \, . \vspace{-0.1cm}
\end{equation}
Here $j_{x,i}(t)$ is the conserved (uniform or staggered) magnetization current
density that flows from the lattice site $x$ to the neighboring lattice site 
$x+a\hat i$ at the time $t$. The continuity equation \eqref{continuity} and
the gradient equation \eqref{gradient} can be derived from the underlying 
real-time path integral that was discussed in detail in \cite{Ban14,Heb15}. The 
corresponding spin configurations together with the resulting values for 
$\rho_x(t)$ and $j_{x,i}(t)$ are illustrated in Fig.~\ref{configs} for the two 
dissipative processes.
 
\begin{figure}[t!]
\includegraphics[width=0.45\textwidth]{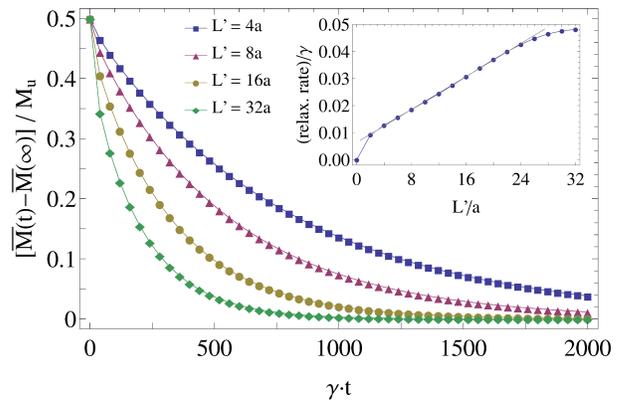}
\caption{[Color online] 
\textit{Real-time evolution of the total uniform magnetization in the first subsystem $\overline{M}$ 
for different values of $L'$ ($L=32a$, total uniform magnetization $M_u=\frac{1}{2}(L/a)^2=512$).
Inset: Late-time relaxation rate as a function of $L'$.}}
\label{magflow}
\end{figure}

We have also investigated the time-dependence of the total uniform magnetization 
in the first subsystem (initially ferromagnetic) as a function of the opening size $L'$ in dissipative process 1 (cf.\ Fig.~\ref{magflow}).
The final state, for which the magnetization is homogeneously distributed throughout the entire system,
is reached exponentially at long times. The relaxation rate then depends linearly on the opening size $L'$ over a 
wide range of values of $L'$.

\begin{figure}[b!]
\includegraphics[width=0.45\textwidth]{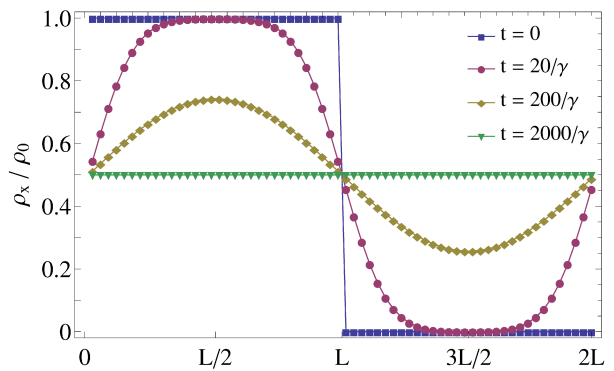}
\caption{[Color online] 
\textit{The $1$-d profile of the uniform magnetization density ($L' = L = 32 a$)
 evolves from a step function at the initial time to a uniform distribution at late times.}}
\label{diffusion1d}
\end{figure}

For the largest possible size of openings, $L' = L$, the diffusion equation 
reduces to a 1-d problem which can even be solved analytically. The resulting
profile of the magnetization density, illustrated in Fig.~\ref{diffusion1d}, is 
given by
\begin{eqnarray}
\rho_x(t)&=&\frac{\rho_0}{2} \sum_{\substack{n=1\\(n\,\text{odd})}}
^{2L/a-1}\frac{a \sin\left(\frac{\pi n}{2L}\left(2x+a\right)\right)}{L \sin\left(\frac{\pi n a}{2L}\right)} \nonumber \\
         &\times&\exp\left(- 2\gamma \sin^2\left(\frac{\pi n a}{2L}\right) t\right) + \frac{\rho_0}{2} \, .
\end{eqnarray}

Certain features of the dissipative processes discussed here resemble classical
physics. For example, if finally all spins are projected along the 3-axis at 
the end of the real-time evolution, the spin configurations on the two branches 
of the Keldysh contour become identical \cite{Ban14,Heb15} and their evolution 
reduces to a Kawasaki dynamics \cite{Kaw66} (cf.\ Fig.~\ref{configs}), 
which can be captured by a classical diffusion equation. Other aspects of the 
same dynamics, including the time-evolution of entanglement, do not have this 
feature, thus underscoring the quantum nature of the corresponding real-time 
processes. In particular, while the expectation value of the conserved quantity 
obeys a classical diffusion equation, its probability distribution can only be 
calculated quantum mechanically. We emphasize that the
presented method is not restricted to probing only diagonal elements of the density
matrix. Most notably, two-point correlation functions reflecting off-diagonal 
entries of the density matrix could also be measured very efficiently via
improved estimators \cite{Bro98}.

It would be most interesting to investigate the non-dissipative pure Hamiltonian
dynamics of large closed quantum systems. Due to very severe complex phase 
problems this is most likely impossible on a classical computer. On the other 
hand, quantum simulators, for example, using ultracold atoms in optical 
lattices, are ideally suited for such investigations. It is conceivable to experimentally
design a dissipative environment which acts as a projector on singlet and triplet states 
(process 1) with current technology \cite{Tro10}, whereas the realization of process 2
is probably more involved. On the other hand, as we have shown, engineered purely dissipative 
processes are accessible to very efficient real-time simulation of large open quantum systems 
using classical computers. Such real-time processes thus provide a bridge between classical and 
quantum simulation. It will be most interesting to explore other processes,
including a weakly coupled Hamiltonian or non-Hermitean Lindblad operators, in 
order to explore the territory connecting classical and quantum 
simulation of quantum dynamics in real time.

\begin{acknowledgments}
We like to thank I.\ Bloch, D.\ B\"odecker, M.\ Di Ventra, and P.\ Zoller for 
illuminating discussions and M.\ Kon for his collaboration on \cite{Ban14}
which forms the basis of the work presented here. The research leading to these 
results has received funding from the Ministry of Science and Technology (MOST) 
of Taiwan under Grant No. 102-2112-M-003-004-MY3, from the Schweizerische Nationalfonds 
zur F\"orderung der Wissenschaftlichen Forschung and from the European Research Council 
under the European Union’s Seventh Framework Programme (FP7/2007-2013)/ERC Grant Agreement 339220.
\end{acknowledgments}


\begin{thebibliography}{99}
 
\bibitem{Tro05}
M.\ Troyer and U.-J.\ Wiese, Phys.\ Rev.\ Lett.\ 94 (2005) 170201.

\bibitem{Bie95}
W. Bietenholz, A.\ Pochinsky, and U.-J.\ Wiese, 
Phys.\ Rev.\ Lett.\ 75 (1995) 4524.
 
\bibitem{Cha99}
S.\ Chandrasekharan and U.-J.\ Wiese, Phys.\ Rev.\ Lett.\ 83 (1999) 3116.

\bibitem{Cha10}
S.\ Chandrasekharan, Phys.\ Rev.\ D82 (2010) 025007.

\bibitem{Cha12}
S.\ Chandrasekharan and A.\ Li, Phys.\ Rev.\ Lett.\ 108 (2012) 140404.

\bibitem{Cha13}
E.\ F.\ Huffman and S.\ Chandrasekharan, Phys.\ Rev.\ B89 (2014) 111101.

\bibitem{Whi92}
S.\ R.\ White, Phys.\ Rev.\ Lett.\ 69 (1992) 2863.

\bibitem{Sch05}
U.\ Schollw\"ock, Rev.\ Mod.\ Phys.\ 77 (2005) 259.

\bibitem{Vid03}
G.\ Vidal, Phys.\ Rev.\ Lett.\ 91 (2003) 147902.

\bibitem{Whi04}
S.\ R.\ White and A.\ E.\ Feiguin, Phys.\ Rev.\ Lett.\ 93 (2004) 076401.

\bibitem{Ver04}
F.\ Verstraete, J.\ J.\ Garcia-Ripoll, and J.\ I.\ Cirac, 
Phys.\ Rev.\ Lett.\ 93 (2004) 207204.

\bibitem{Zwo04}
M.\ Zwolak and G.\ Vidal, Phys.\ Rev.\ Lett.\ 93 (2004) 207205.

\bibitem{Dal04}
A.\ J.\ Daley, C.\ Kollath, U.\ Schollw\"ock, and G.\ Vidal, 
J.\ Stat.\ Mech.: Theor.\ Exp.\ P04005 (2004).

\bibitem{Bar09}
T.\ Barthel, U.\ Schollw\"ock, and S.\ R.\ White, Phys.\ Rev.\ B79 (2009)
245101.

\bibitem{Piz13}
I.\ Pizorn, V.\ Eisler, S.\ Andergassen, and M.\ Troyer, 
New\ J.\ Phys.\ 16 (2014) 073007.

\bibitem{Cor74}
J.\ M.\ Cornwall, R.\ Jackiw, and E.\ Tomboulis, Phys.\ Rev.\ D10 (1974) 2428.

\bibitem{Ber02}
J.\ Berges, Nucl.\ Phys.\ A699 (2002) 847.

\bibitem{Aar02}
G.\ Aarts, D.~Ahrensmeier, R.~Baier, J.~Berges, and J.~Serreau, 
Phys.\ Rev.\ D66 (2002) 045008.

\bibitem{Ber08}	
J.\ Berges, A.\ Rothkopf, and J.\ Schmidt, 
Phys.\ Rev.\ Lett.\ 101 (2008) 041603.

\bibitem{Die08}
S.\ Diehl, A.\ Micheli, A.\ Kantian, B.\ Kraus, H.\ P.\ B\"uchler, and P.\ Zoller,
Nature Phys.\ 4 (2008) 878.
 
\bibitem{Dal09}
E.\ G.\ Dalla Torre, E.\ Demler, T.\ Giamarchi, and E.\ Altman, 
Nature Phys.\ 6 (2009) 806.

\bibitem{Die10}
S.\ Diehl, A.\ Tomadin, A.\ Micheli, R.\ Fazio, and P.\ Zoller, 
Phys.\ Rev.\ Lett.\ 105 (2010) 015702.

\bibitem{Mue12}
M.\ M\"uller, S.\ Diehl, G.\ Pupillo, and P.\ Zoller, 
Adv.\ Atom.\ Mol.\ Opt.\ Phys.\ 61 (2012) 1.

\bibitem{Sie13}
L.\ M.\ Sieberer, S.\ D.\ Huber, E.\ Altman, and S.\ Diehl, 
Phys.\ Rev.\ Lett.\ 110 (2013) 195301.

\bibitem{DeG13}
C.\ De Grandi, A.\ Polkovnikov, and A.\ W.\ Sandvik, 
J.\ Phys.: Cond.\ Matt.\ 25 (2013) 404216.

\bibitem{Hor13}
B.\ Horstmann, J.\ I.\ Cirac, and G.\ Giedke, Phys.\ Rev.\ A87 (2013) 012108.

\bibitem{Chi15}
C.-C.\ Chien, S.\ Peotta, and M.\ Di Ventra, arXiv:1504.02907.

\bibitem{Ban14}
D.\ Banerjee, F.-J.\ Jiang, M.\ Kon, and U.-J.\ Wiese, 
Phys.\ Rev.\ B90 (2014) 241104.

\bibitem{Heb15}
F.\ Hebenstreit, D.\ Banerjee, M.\ Hornung, F.-J.\ Jiang, F.\ Schranz, and 
U.-J.\ Wiese, Phys.\ Rev.\ B92 (2015) 035116.

\bibitem{Rau01}
R.\ Raussendorf and H.\ J.\ Briegel, Phys.\ Rev.\ Lett.\ 86 (2001) 5188.

\bibitem{Nie01}
M.\ A.\ Nielsen, Phys.\ Lett.\ A308 (2003) 96.

\bibitem{Chi02}
A.\ M.\ Childs, D.\ Deotto, E.\ Farhi, J.\ Goldstone, S.\ Gutmann, and A.\ J.\ Landahl, 
Phys.\ Rev.\ A66 (2002) 032314.

\bibitem{Ali04}
P.\ Aliferis and D.\ W.\ Leung, Phys.\ Rev.\ A70 (2004) 062314.

\bibitem{Ver09}
F.\ Verstraete, M.\ M.\ Wolf, and J.\ I.\ Cirac, Nature Phys.\ 5 (2009) 633.

\bibitem{Kra11}
H.\ Krauter, C.\ A.\ Muschik, K.\ Jensen, W.\ Wasilewski, J.\ M.\ Petersen, J.\ I.\ Cirac, and E.\ S.\ Polzik,
Phys.\ Rev.\ Lett.\ 107 (2011) 080503.

\bibitem{Sug05}
M.\ Sugawara, J.\ Chem.\ Phys.\ 123 (2005) 204115.

\bibitem{Pec06}
A.\ Pechen, N.\ Il'in, F.\ Shuang, and H.\ Rabitz, Phys.\ Rev.\ A74 (2006)
052102.

\bibitem{Cir12}
J.\ I.\ Cirac and P.\ Zoller, Nature Phys.\ 8 (2012) 264.

\bibitem{Lew12}
M.\ Lewenstein, A.\ Sanpera, and V.\ Ahufinger, ``Ultracold Atoms in Optical 
Lattices: Simulating Quantum Many-Body Systems'', Oxford University Press 
(2012).

\bibitem{Blo12} 
I.\ Bloch, J.\ Dalibard, and S.\ Nascimbene, Nature Phys.\ 8 (2012) 267.

\bibitem{Bla12}
R.\ Blatt and C.\ F.\ Ross, Nature Phys.\ 8 (2012) 277.

\bibitem{Kos72}
A.\ Kossakowski, Rep.\ Math.\ Phys.\ 3 (1972) 247.

\bibitem{Lin76}
G.\ Lindblad, Commun.\ Math.\ Phys.\ 48 (1976) 119.

\bibitem{Kra83}
K.\ Kraus, States, Effects and Operations, Fundamental Notions of Quantum 
Theory, Academic, Berlin (1983).

\bibitem{Sch61}
J.\ Schwinger, J.\ Math.\ Phys.\ 2 (1961) 407.

\bibitem{Kel65}
L.\ V.\ Keldysh, Sov.\ Phys.\ JETP 20 (1965) 1018.

\bibitem{Eve93}
H.\ G.\ Evertz, G.\ Lana, and M.\ Marcu, Phys.\ Rev.\ Lett.\ 70 (1993) 875.

\bibitem{Wie94}
U.-J.\ Wiese and H.-P.\ Ying, Z.\ Phys.\ B93 (1994) 147.

\bibitem{Bea96}
B.\ B.\ Beard and U.-J.\ Wiese, Phys.\ Rev.\ Lett.\ 77 (1996) 5130.

\bibitem{Ger09}
U.\ Gerber, C.\ P.\ Hofmann, F.-J.\ Jiang, M.\ Nyfeler, and U.-J.\ Wiese,
J.\ Stat.\ Mech.: Theor.\ Exp.\ P03021 (2009).

\bibitem{Ger11}
U.\ Gerber, C.\ P.\ Hofmann, F.-J.\ Jiang, G.\ Palma, P.\ Stebler, and U.-J.\ Wiese,
J.\ Stat.\ Mech.: Theor.\ Exp.\ P06002 (2011).

\bibitem{Kaw66}
K.\ Kawasaki, Phys.\ Rev.\ 145 (1966) 224.

\bibitem{Bro98}
R.\ Brower, S.\ Chandrasekharan, and U.-J.\ Wiese, Physica\ A261 (1998) 520.

\bibitem{Tro10}
S.\ Trotzky, Y.-A.\ Chen, U.\ Schnorrberger, P.\ Cheinet, and I.\ Bloch,
Phys.\ Rev.\ Lett.\ 105 (2010) 265303.

\end{thebibliography}
\end{document}